\documentclass{article}
\usepackage{ijcai21}
\usepackage{appendix}
\usepackage[inline]{enumitem}
\usepackage{color, colortbl}
\definecolor{Gray}{gray}{0.9}
\usepackage[first=0,last=9]{lcg}

\usepackage{times}

\usepackage{soul}
\usepackage{url}
\usepackage[hidelinks]{hyperref}
\usepackage[utf8]{inputenc}
\usepackage[small]{caption}
\usepackage{graphicx}
\usepackage{amsmath}
\usepackage{booktabs}
\usepackage{xcolor}
\usepackage{latexsym}
\usepackage[utf8]{inputenc}
\makeatletter
\newcommand*\titleheader[1]{\gdef\@titleheader{#1}}
\AtBeginDocument{%
\let\st@red@title\@title
\def\@title{%
\bgroup\normalfont\large\centering\@titleheader\par\egroup
\vskip1.5em\st@red@title}
}
\makeatother

\title{Decoys in Cybersecurity: An Exploratory Study to Test the Effectiveness of 2-sided Deception}

\titleheader{\emph{IJCAI-21 1\textsuperscript{st} International Workshop on Adaptive Cyber Defense}}

\author{
Palvi Aggarwal\footnote{Contact Author}\and
Yinuo Du\and
Kuldeep Singh\And
Cleotilde Gonzalez\\
\affiliations
Carnegie Mellon University, USA\\
\emails
\{palvia, yinuod, kuldeep2\}@andrew.cmu.edu,
coty@cmu.edu
}

\begin{document}

\maketitle

\begin{abstract}
One of the widely used cyber deception techniques is decoying, where defenders create fictitious machines (i.e., honeypots) to lure attackers. Honeypots are deployed to entice attackers, but their effectiveness depends on their configuration as that would influence whether attackers will judge them as ``real" machines or not. In this work, we study two-sided deception, where we manipulate the observed configuration of both honeypots and real machines. The idea is to improve cyberdefense by either making honeypots ``look like'' real machines or by making real machines ``look like honeypots.'' We identify the modifiable features of both real machines and honeypots, and conceal these features to different degrees. In an experiment, we study three conditions: default features on both honeypot and real machines, concealed honeypots only, and concealed both honeypots and real machines. We use a network with 40 machines where 20 of them are honeypots. We manipulate the features of the machines, and using an experimental testbed (HackIT), we test the effectiveness of the decoying strategies against humans attackers. Results indicate that: Any of the two forms of deception (conceal honeypots and conceal both honeypots and real machines) is better than no deception at all. We observe that attackers attempted more exploits on honeypots and exfiltrated more data from honeypots in the two forms of deception conditions. However, the attacks on honeypots and data exfiltration were not different within the deception conditions. Results inform cybersecurity defenders on how to  manipulate the observable features of honeypots and real machines to create uncertainty for attackers and improve cyberdefense.
\end{abstract}

\section{Introduction} 
As we are increasing the number of devices available in the network, our cyber space is continuously getting broader, bigger, and dramatically intensifying the potential attack surface for adversaries. Currently, most data are digitized and stored in the organizations’ servers, making them a valuable target for attackers. Advanced persistent threats (APT), social engineering attacks, insider attacks, and other forms of attacks are becoming more common. Frequent strategies used in data breaches include Ransomware, Malware, Phishing, and application vulnerabilities \cite{verizon}. In 2020, more than 37 billion data records were compromised with increase of 141\% from year 2019 \cite{security}. According to Ponemon Institute report \cite{ibm}, in the year 2020, on average, companies required 280 days (207 days to identify and 73 days to contain) a breach. These numbers are only discussing the attacks that were discovered. It is likely that the reality is significantly worse as there are unreported and undiscovered attacks. These findings show that the organizations security capabilities are not sufficient to address current cyber threats. 

According to Sun Tzu, “All warfare is based on deception." Deception, providing false information to the opponent, is a powerful strategy for cyberdefense \cite{gonzalez2020design,jones1989reflections}. By the strategic use of deceptive techniques, system defenders can mislead and confuse attackers, enhancing their defensive capabilities over time. Common deception strategies used in cyberdefense include: signaling, masking, and decoying \cite{gonzalez2020design,schlenker2018deceiving,mcqueen2009deception,whaley1982toward}. Signaling strategically reveals information about a defensive strategy to the attacker to influence the attacker's decision making\cite{xu2016signaling,gonzalez2020design,cranford2018learning}. Masking techniques make a real object undetectable to hide information behind benign programs (e.g., hiding information behind an image in an email message \cite{aggarwal2020exploratory}), while decoying presents a false object (i.e., honeypot) to grab attention by showcasing fake but relevant information \cite{spitzner2003honeypots}.

Honeypot is perhaps the most common decoying strategy in cybersecurity. Honeypots mimic the configuration of real machines, which might be the attacker's target. Honeypots are used for detection to catch illicit interactions or in prevention, to assist in slowing adversaries down \cite{almeshekah2015using,almeshekah2016cyber}. The effectiveness of honeypots depends on several factors including the configuration of honeypot features, the allocation of honeypots in the network, and the number of honeypots in the network. However, to make honeypots an effective cyber deception technique, they must be implemented in such a way that they appear as real machines to cyber attackers. That is, one needs not only a solid technical implementation behind honeypots but also a solid understanding of human adversarial behavior. For proper implementation of honeypots, defenders need to carefully design the level of deception that would most successfully exploit the trust of attackers. Previous research has manipulated the amount and timing of deception, type and frequency of signals, and the features of the honeypot to exploit the attacker's beliefs \cite{aggarwal2016cyber,miah2020concealing,cranford2020adaptive}. However, past research considered the abstract features of honeypots and has not evaluated the algorithms against human attackers. 
 
The open-source nature of honeypot software that most defenders use to create honeypots makes it easy for adversaries to identify honeypot features (i.e., attackers also have access to such open source software). Deploying honeypots with these features can hurt security rather than help. Moreover, common honeypots are designed so that they look exactly like the real machine when probed by an adversary or they are completely different. However, this might not be necessary as suggested by Miah \shortcite{miah2020concealing}. Instead of manipulating the features of a honeypot to make it look like a real machine, it is possible to use 2-sided deception i.e., make honeypots look like real machines and make real machines look like honeypots. However, the effectiveness of 2-sided deception has not been tested with human adversaries. In this paper, we advance past research by experimentally testing the effectiveness of 2-sided deception. Using an experimental testbed, HackIT, we evaluate the default configuration of honeypots and real nodes against concealed honeypot (1-sided deception, i.e., making honeypots identical to real machines) and concealed honeypots and real nodes (2-sided deception i.e., concealing the configuration of both honeypots and real nodes). 


\section{Background}
Honeypots have been used as an effective tool for cyberdeception since their inception \cite{spitzner2003honeypots}. Honeypots have been used as a decoy machine in the network to gain an attacker's attention. For example, honeypots are often configured with exploitable vulnerabilities so that attackers could easily attack such machines. Honeypots are also used as a tool to gather the behavioral patterns of malicious attackers. To enhance the effectiveness of honeypots, research has focused on various factors such as the location of honeypots in the network, network size and topology, and the configuration.

Prior research has developed game-theoretic algorithms to strategically allocate honeypots in the network \cite{bilinski2018optimal,aggarwalhoneypot2021,anwar2020honeypot}. Bilinski \shortcite{bilinski2018optimal} proposed a honeynet game to optimally allocate honeypots in the network to secure high-valued resources. Anwar~\shortcite{anwar2020honeypot} proposed a scalable honeypot allocation algorithm with the importance of different machines in the network, attackers and defenders cost-effectiveness, and attackers' information level considered.  

Furthermore, research has evaluated the influence of network size and network topology on adversary’s cyberattack decisions \cite{katakwar2020influence,achleitner2017deceiving}. The topology of the defended network in which honeypots are deployed also matters, especially on reconnaissance missions. Achleitner~\shortcite{achleitner2017deceiving} proposed a reconnaissance deceptive virtual network topology that can be generated based on Software Defined Network (SDN) and showed its potential to delay malicious network scans. 

Research effort has also been devoted to determine the optimal configuration of honeypots. Through masking the systems’ attributes to disguise valuable
information, it is possible to increase the attacker’s time spent in planning and compromising the network. Shi~\shortcite{shi2020learning} proposed a solution to utilize the \textit{feature deception problem(FDP)} model and to leverage attacker preferences known by defenders. Aggarwal~\shortcite{aggarwal2020exploratory} evaluated optimal masking strategies in comparison with the random masking strategy against human adversaries. However, there is a common assumption in the literature that either the honeypots look exactly like the real machine when probed by an adversary or they are completely different. Thus, the majority of the past research has focused on manipulating the features of the honeypot to make them identical to real machines \cite{shi2016game}. This might not be true. There are several side channel techniques to detect low-interaction honeypots. Even in the case of high-interaction honeypots, if the machine is probed from within the host (i.e., after being compromised), there may be certain indicators (e.g., no user activity, no network traffic) that may reveal that the machine is a decoy. 

Assuming that experienced attackers would detect honeypots, a few researchers have studied the concept of "fake honeypot" i.e., creating honeypot looking systems in the network \cite{rowe2007defending,shi2016game,miah2020concealing}. Most recent work by Miah \shortcite{miah2020concealing} introduced a game-theoretic model of two-sided deception that use of both fake-honeypots and fake-real systems in the network (2-sided deception). Instead of solely making honeypots look like real systems (\textit{concealed-honeypot}), a defender can modify some features of the real machines and make them look like honeypots (\textit{concealed-real}). Miah \shortcite{miah2020concealing} numerically evaluated the 2-sided deception model and thus, how human attackers would make decisions with such models is still unknown. 

Evaluation of these game theoretic algorithms of cyberdeception is comparatively rare. Ferguson \shortcite{ferguson2021examining,ferguson2018tularosa} studied attacker's behavior against cyber deception using experiments. A couple of platforms have been used to experiment with cyber deception and dynamic honeypots. For example, CyberVAN, HackIT, and Cybersecurity Deception Experimentation System (CDES) \cite{acosta2020cybersecurity,aggarwal2019hackit,chadha2016cybervan}. In this paper, we will evaluate the effectiveness of cyber deception using a tool called HackIT, i.e., a testbed to deploy deception and conduct experiments to test different levels of deception using honeypots against attackers \cite{aggarwal2019hackit}.

\begin{table*}[!htb]
\renewcommand{\arraystretch}{1.7}
\caption{\label{tab:feature}The configuration of Real and Honeypot machines}
\scalebox{0.80}{
\begin{tabular}{p{4cm}|p{4.1cm}|p{3.9cm}|p{3.9cm}|p{4.1cm}}

\hline
\rowcolor{Gray}
\textbf{Features}                                     & \textbf{Default Honeypot}                                                                                                           & \textbf{Concealed-Honeypot}        & \textbf{Default Real Machine}    & \textbf{Concealed-Real}                                                                                                             \\ \hline
OS                                                    & Obsolete OS                                                                                                                         & Up to date                    & Up to date                    & Obsolete OS                                                                                                                           \\ \hline
Ports \& Services                                     & \begin{tabular}[c]{@{}l@{}}Normal Ports, Honeypot Ports\end{tabular} & Normal Ports       & Normal Ports                  & \begin{tabular}[c]{@{}l@{}}Normal Ports, Honeypot Ports\end{tabular} \\ \hline
Exploit Info                                          & Obsolete                                                                                                                            & Up to date                    & Up to date                    & Obsolete                                                                                                                            \\ \hline
Exploit Success Rate                                  & 100\%                                                                                                                               & 40\%                          & 40\%                          & 80\%                                                                                                                                \\ \hline
Link Latency                                          & 1ms                                                                                                                                 & 0.2ms                         & 0.2ms                         & 1ms                                                                                                                                 \\ \hline
Virtual/Physical Machine                              & VM                                                                                                                                  & Physical                      & Physical                      & VM                                                                                                                                  \\ \hline
Running Processes                                     & 2 processes                                                                                                                 & 10 processes          & 10 processes          & 2 processes                                                                                                                 \\ \hline
File System (1 user folder with 5$\sim$6 sub-folders) & 4$\sim$5 empty or access-deny folders                                                                                               & 1 empty or access-deny folder & 1 empty or access-deny folder & 4$\sim$5 empty or access-deny folders                                                                                               \\ \hline
\end{tabular}}

\end{table*}
\section{Human Experiment} 
We conduct an exploratory study using an experimental testbed called HackIT \cite{aggarwal2019hackit}. HackIT provides a capability to create a network of machines, configure the observable features of machines to deploy deception, and manipulate network size and topology for conducting human-in-the-loop experiments. In this paper, we use HackIT to create a network of 40 machines and deploy deception using different configurations of 20 honeypots and 20 real machines.
\subsection{Features of Honeypot and Real Machines}
Attackers continuously gain information about the network during different cyber attack phases (reconnaissance - exploit - post exploit). To acquire the information, attackers use network scanning tools, specialized penetration tools, and advanced machine learning tools. These tools have different degrees of reliability and are often used in combination.

Defenders often deploy honeypots in the network to divert attackers from real machines to honeypots. Most of the time, the default configuration of the honeypot is well known to the attackers and is not effective when deployed. Here we studied the default configurations of honeypots and their weaknesses (Honeyd, Dionaea). We also researched the features of real machines that are modifiable. We selected 8 features, i.e., Operating System (OS), ports and services, exploit information, exploit success rate, link latency, virtual or physical machines, number of running processes, and the file system. These features, listed in Table \ref{tab:feature}, were used to configure the observable features of the machines used in our study. 

A default honeypot is a machine with all commonly used features and indicators. Feature assignment listed in table \ref{tab:feature} column \textit{default honeypot} comes from the default configuration of open source honeypot software such as Cowrie, Dionaea \textit{etc.} \cite{cabral2019review}. A typical incautiously deployed honeypot appears to be running an obsolete operating system and have some suspicious and default ports opened \cite{spangler2003analysis,grimes2006honeypots,dahbul2017enhancing}. The default configuration of honeypots also includes exploitable vulnerabilities which are usually patched on real host, abnormal link latency (round trip time), absence of necessary running processes, and empty and inaccessible directories of file system \cite{fu2006recognizing,hayatle2012dempster}. Virtual/sand boxed environment is another major source of flaws \cite{chen2008towards,papazis2019detecting} that inferior honeypots fail to take care of. We select the \textit{Default Real Machine} from real working systems as listed in table \ref{tab:feature}. 

Concealed-honeypots are disguised with the real machine's appearance, i.e., we modified the features of the honeypots to make them look like real machines. The selected features are verified to be either configurable or hide-able on honeypots \cite{tsikerdekis2018approaches}. Similarly, Concealed-Real machines are configured such as they look like honeypots through attack surface manipulation techniques \cite{wang2010honeypot,albanese2016deceiving}.

\subsection{Experiment Design}

In this experiment, we study the effectiveness of two-sided deception strategy. We define three between-subject conditions: \textit{default} (both real and honeypots have default configurations), \textit{concealed-honeypots} (only honeypot's configuration is concealed) and \textit{concealed-real-and-honeypot} (both honeypot and real machines are concealed). Each condition consists of two identical rounds. The proportion of honeypots (20 out of 40 machines) remains the same across conditions and rounds. The honeypots were allocated randomly in the network. 

\subsection{HackIT Task}
In this experiment, the participant played the role of an attacker. The goal of an attacker was to steal confidential files called "pin.txt" containing credit card pin numbers; from different web server systems on the network. There were 40 systems in the network and the participant's can exploit as many systems as they can in the network within a given time. For the exploration and exploitation of a system on the network, the participants were provided tools which were accessible through commands. The commands for exploration and exploitation of systems are: \textit{(1) nmap -sL all, (2) nmap System (3) nmap System [-rtt], (4) checkHS, (5) info\_ exploit exploit, (6) use\_exploit}. 

\begin{figure}[!htb]
\begin{center}
\includegraphics[scale=0.50]{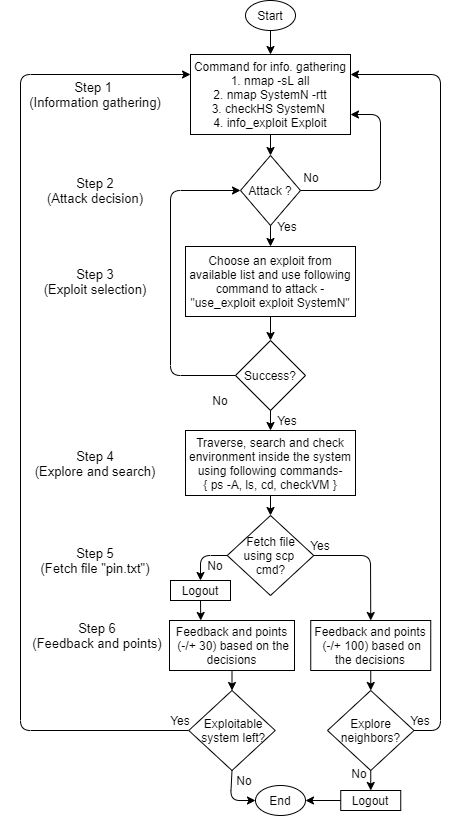}
\caption{The steps involved in repetitive round 1 and round 2}
\label{fig:HackITTask}
\end{center}
\end{figure}

Figure \ref{fig:HackITTask} describes the various steps involved in the task. The first step in the task is information gathering. As shown in Figure \ref{fig:HackITTask}, attackers could use nmap, checkHS, and info\_ exploit commands for the information gathering phase. The \textit{“nmap -sL all”} command is to find what web servers are available to probe. To know about the list of ports, vulnerabilities present on the corresponding ports, and the operating system \textit{“nmap System”} command can be used. The \textit{“nmap System [-rtt]”} command provides the benchmark round-trip-time of production systems and the RTT of System. To use RTT command participants have to pay 10 seconds and this time is deducted from the experiment remaining time of participant. The \textit{“checkHS”} command is to detect the existence of deceptive strategies. The reliability of \textit{“checkHS”} command is only 50\% and it cost approximately 10 seconds to participants. It generate a score from 0 to 1 representing the likelihood of the system being deceptive (0 = “not deceptive” to 1=“deceptive”). To find the most promising attack path, participants can collect more detailed information, including its disclosure date, about the presented exploits using \textit{“info\_exploit exploit”}. Attackers could decide to attack the system they explored or gather more information about other systems (step 2, Figure \ref{fig:HackITTask}).

Once the participants are ready to exploit a machine, they need to select an exploit as defined in step 3 of Figure \ref{fig:HackITTask}. To exploit a vulnerability on a server, they need to execute the \textit{“use\_ exploit”} command. The use\_exploit command accepts two parameters: the vulnerability to be attacked and the target system name. For example, the participant can run the following command to exploit sql injection vulnerability on System1: \textit{"use\_exploit sql\_injection System1"}.The “use\_exploit” command might require multiple tries to gain success. The participants can also switch to another vulnerability on the same system or another system. 

Once the participants gain access to the server, they may do further reconnaissance with the following commands- {`ls',`cd', `ps -A', and `checkVM'} as shown in step 4 of Figure \ref{fig:HackITTask}. The \textit{``ls''} command lists the files present in the current directory. The \textit{``cd''} command helps to traverse through the directories. The \textit{``ps -A''} command lists all running processes on the server. The \textit{``checkVM''} command indicate whether the machine is running on a physical system or in a virtual environment. After collecting enough information and traversing through various directories, participants may decide whether to steal the file or leave the system without stealing the file (step 5, Figure \ref{fig:HackITTask}). If they decide to steal the file, they may use the scp command to transfer the “pin.txt” file from the server to their system. The scp command takes two parameter filename and destination and works in the following way: \textit{scp pin.txt address}. The participant gets feedback about the success of the operation and the points earned or lost in the task (step 6, Figure \ref{fig:HackITTask}). If participants logout without fetching the file, they gain 30 points if the exploited system was real or lose 30 points if the exploited machine was a honeypot. If participants fetched the file, the rewards and losses were higher. The participant earns 100 points if they exploit a real system, if they exploit a honeypot then they lose 100 points. After that, the participant's can also choose to logout and terminate the attack and then go on to probe, exploit, and steal files from other servers. Participants were provided 60 minutes to complete the task. The task ends once the participant runs out of the allowed time for the task or explores all systems in the network. 

\subsection{Participants}
Participants were recruited through advertisements via various university email groups, social media, cybersecurity targeted groups, and Amazon Mechanical Turk. To be qualified to participate in the study, participants were required to pass an online screening test of basic cybersecurity knowledge, which included questions about network scanning, various types of attacks, etc. The question pool contains 18 questions in total, and each participant was assigned 10 questions randomly. The participant was required to answer 7 questions correctly. A total of 95 participated in the screening test and 71 passed with an average score of 8.4.

Qualified participants were scheduled for an online study of 60 minutes. After qualifying the screening test on cybersecurity knowledge, 48 participants proceeded to the main study. Participants were randomly assigned to one of the three conditions, default (n=14), concealed-honeypot condition (n=13), and concealed-honeypot-real condition (n=21). Participants aged between 18 and 62 years (Mean: 33, SD: 9.9). The education and age demographics information is presented in Table \ref{tab:demo}. After the successful completion of the experiment, all participants were given a base payment of \$12 and could earn a bonus up to \$10. The average time taken to complete this experiment was 60 minutes.

\begin{table}[]
\renewcommand{\arraystretch}{1.1} 
    \centering
     \caption{Demographics}
    \begin{tabular}{l|c|r}
    \hline
    \rowcolor{Gray}
        \textbf{ Category}&\textbf{Sub-category}&\textbf{Percentage}  \\
    \hline
         Sex & Male & 81\% \\
          & Female & 13\% \\
          & Others & 6\% \\
          
    \hline
         Education & Master's  & 33.3\% \\
          & Bachelor's & 60.40\% \\
          & PhD & 6.30\% \\
    \hline
         Degree & Computer Science  & 79.15\% \\
          & Electrical Engineering & 6.25\% \\
           & STEM & 12.50\% \\
          & Other & 4.16\% \\
    \hline
        Cybersecurity Course & Yes  & 77\% \\
          & No & 23\% \\
    \hline
    \end{tabular}
   
    \label{tab:demo}
\end{table}

\subsection{Procedure}

First, participants provided informed consent and filled the demographic information. Next, an instruction video were presented to them followed by text instructions regarding the goal of the task and the general procedure (presented in Appendix A). Instructions were followed with a brief quiz to evaluate their comprehension of the instructions. Participants received feedback if they incorrectly answered a question in the quiz. Once all their questions were clarified, they were allowed to proceed with the experiment. Participants were not informed about the proportion of honeypots in the network. 

The participants were involved in a practice round before entering into the main section of the study. The practice round makes the participant familiar with the tool and provides knowledge about the next main task in the study. In the main repetitive section of the task, the participants have to use different commands to gather information about the systems on the network. Based on the information, the participants have to decide regarding to exploit a system or not. And if they decide to exploit and enter into a system, then they also have to make another decision about stealing a confidential file from the system and transfer it to their system. After the completion of the main task in the experiment, a feedback questionnaire was presented to the participants, where participants provided their feedback about the task and the strategies used by them during the task. 

\section{Results}
We analyzed the attackers actions in the main task. Specifically, we explored the distribution of attacks, the success of deception by measuring the honeypot exploitation rate, data exfiltration rate from honeypots, and the average score earned in each experimental condition.

\subsection{Distribution of System Exploitation}
\begin{figure}[!htb]
\begin{center}
\includegraphics[scale=0.40]{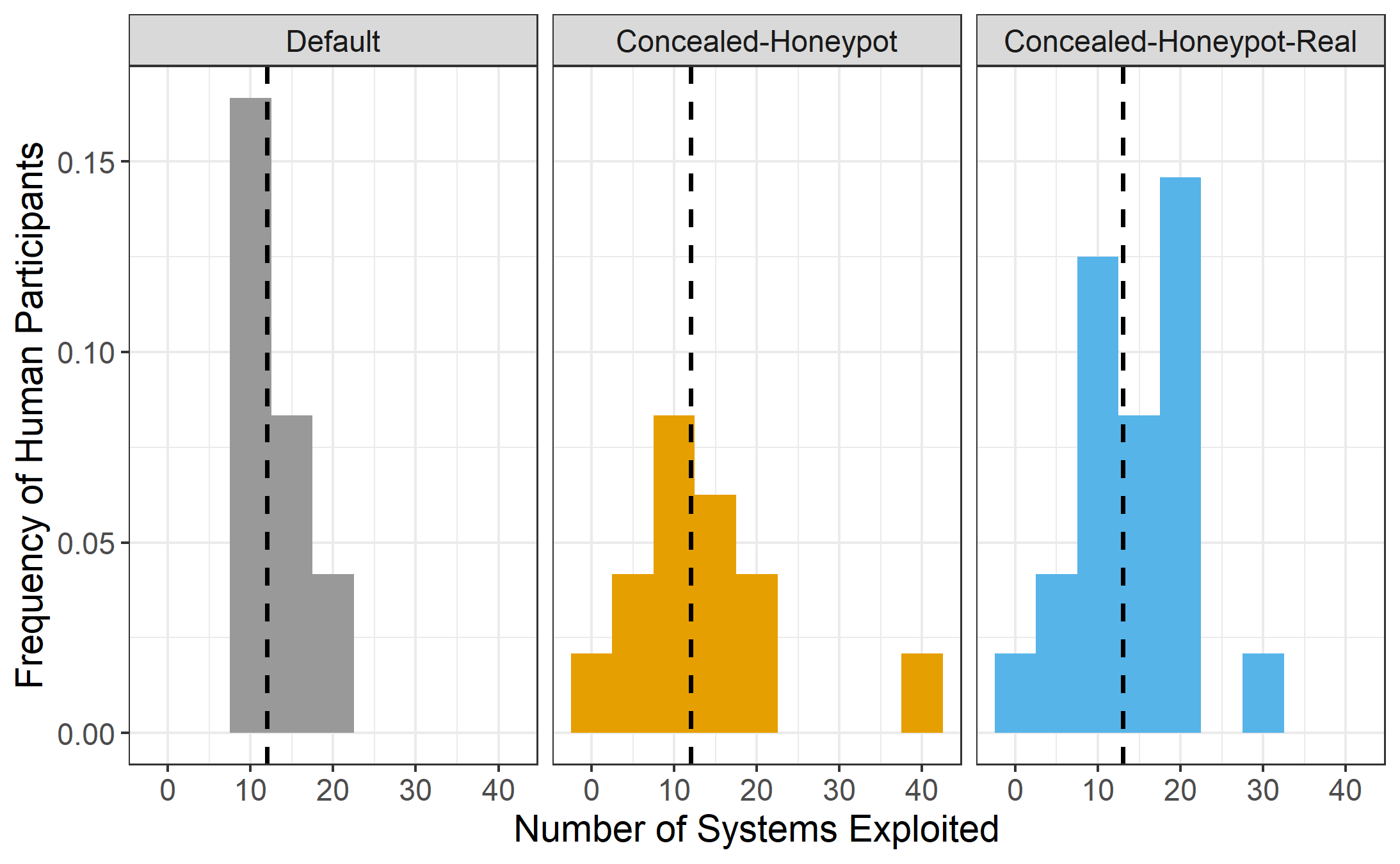}
\caption{Distribution of number of systems exploited in each condition}
\label{fig:ExploitCount}
\end{center}
\end{figure}

To investigate the differences in the three conditions, we analyzed the distribution of the number of systems exploited in each condition in Figure~\ref{fig:ExploitCount}. The median number of systems exploited were similar in three conditions, i.e., 12 systems in the default condition, 12 systems in the concealed-honeypots and 13 systems in the concealed-honeypot-real conditions. Thus, the condition does not have an impact on the median exploitation behavior. The fisher's exact test of distribution indicates the significant difference in the distribution of the three conditions \textit{(p$<$0.05}). 

\subsection{Proportion of Honeypot Attacks}
After the reconnaissance, the participant could decide to exploit the system which could be a honeypot or real. The success of deception could be measured based on the number of attacks on the honeypots. In our manipulation, the attempted exploit may sometimes fail and result in a failed attack. 
We analyzed the proportion of failed honeypot attacks and the proportion of successful honeypot attacks. The failed attacks contributed to the wasting time and resources of attackers. 

\begin{figure}[!htb]
\begin{center}
\includegraphics[scale=0.40]{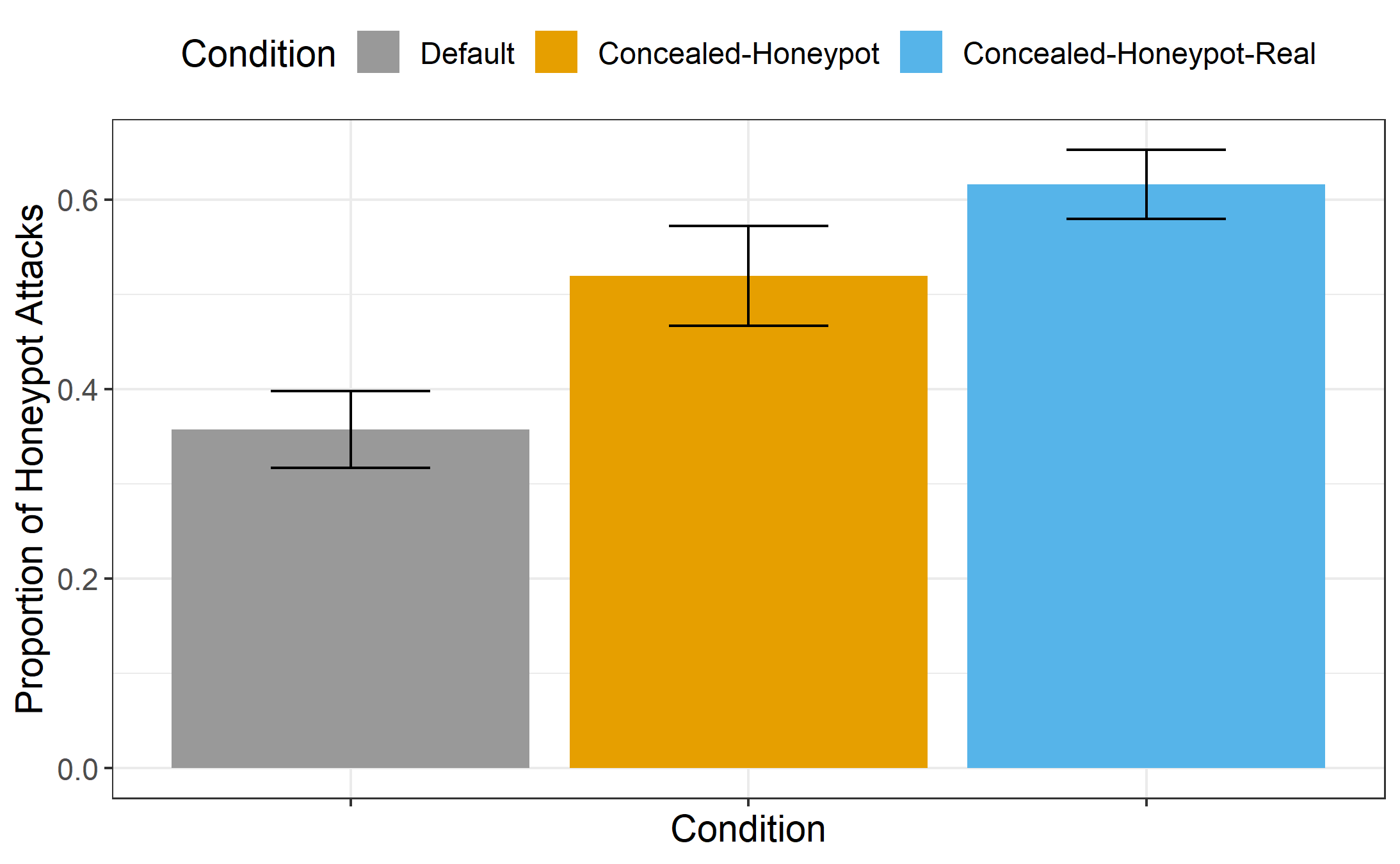}
\caption{Proportion of Honeypot Attacks}
\label{fig:allhoneypot}
\end{center}
\end{figure}

We calculated the proportion of honeypot attacks as shown in Figure \ref{fig:allhoneypot}. There is a statistically significant difference between groups as determined by one-way ANOVA (\textit{F}(2,45) = 9.75, \textit{p} = 0.0001). A Tukey post hoc test revealed that the proportion of honeypot attacks are significantly higher in concealed-honeypot condition(\textit{p} = 0.04) and concealed-Honeypot-Real (\textit{p} = 0.0001) condition compared to the default condition; although there is no statistically significant difference between the concealed-honeypot and concealed-honeypot-real (\textit{p} = 0.25). Our results suggest that any of the two types of deception are better at increasing the attacks on honeypots compared to using no deception at all. Additionally, concealing both real and honeypot machines produces more attacks than only concealing honeypots, although the average difference in the proportion of attacks was not statistically different.

\begin{figure}[!htb]
\begin{center}
\includegraphics[scale=0.40]{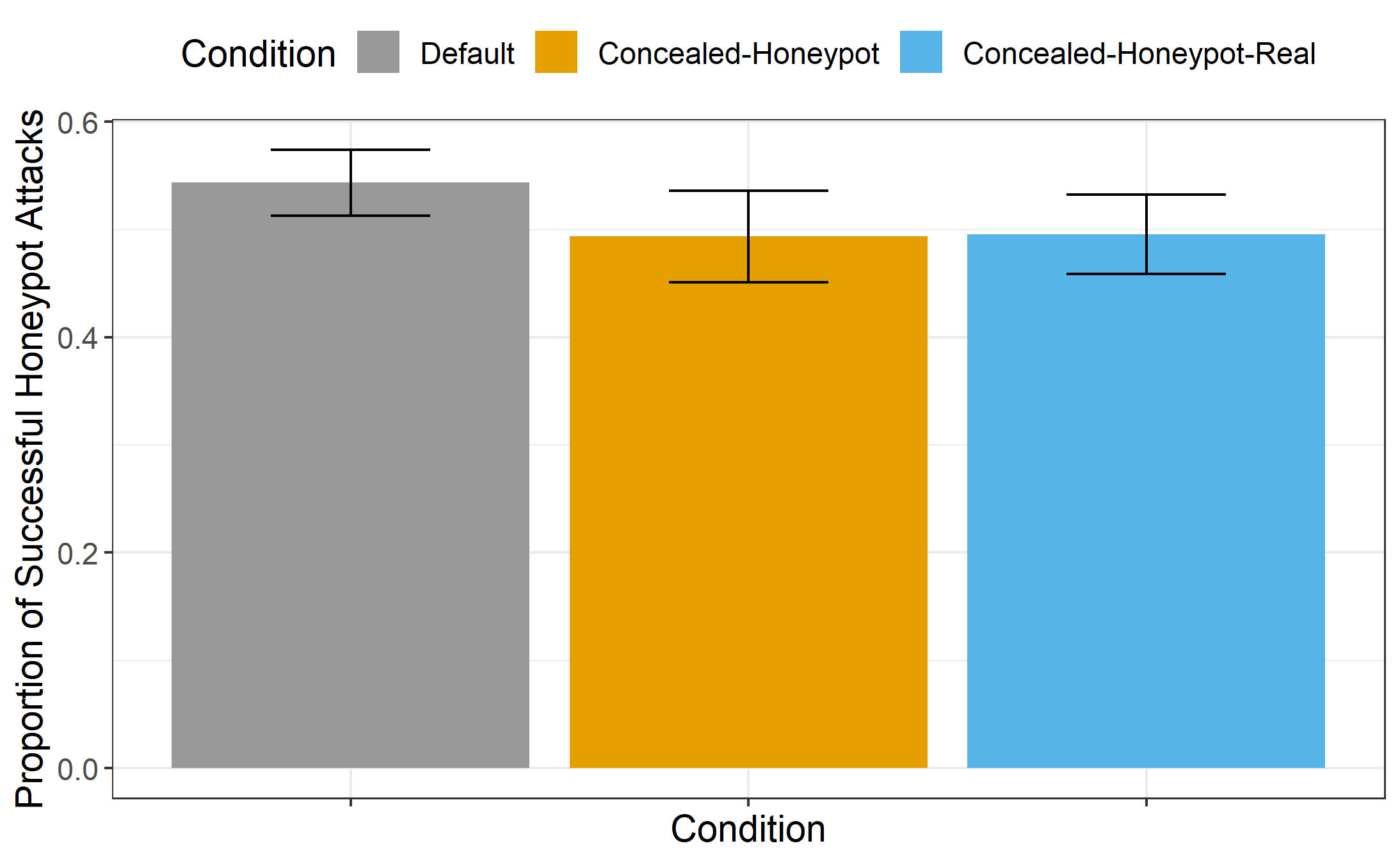}
\caption{Proportion of Successful Honeypot Attacks}
\label{fig:honeypot_attack_proportion}
\end{center}
\end{figure}

Next, we calculated the proportion of successful honeypot attacks in Figure \ref{fig:honeypot_attack_proportion}. We observe that there was no significant difference between conditions as determined by one-way ANOVA (\textit{F}(2,45) = 0.523, \textit{p} = 0.596). Although the deception results in more exploitation attempts, the overall successful attacks on honeypots are similar in the three conditions. 

\subsection{Proportion of Data Exfiltration}
\begin{figure}[!htb]
\begin{center}
\includegraphics[scale=0.40]{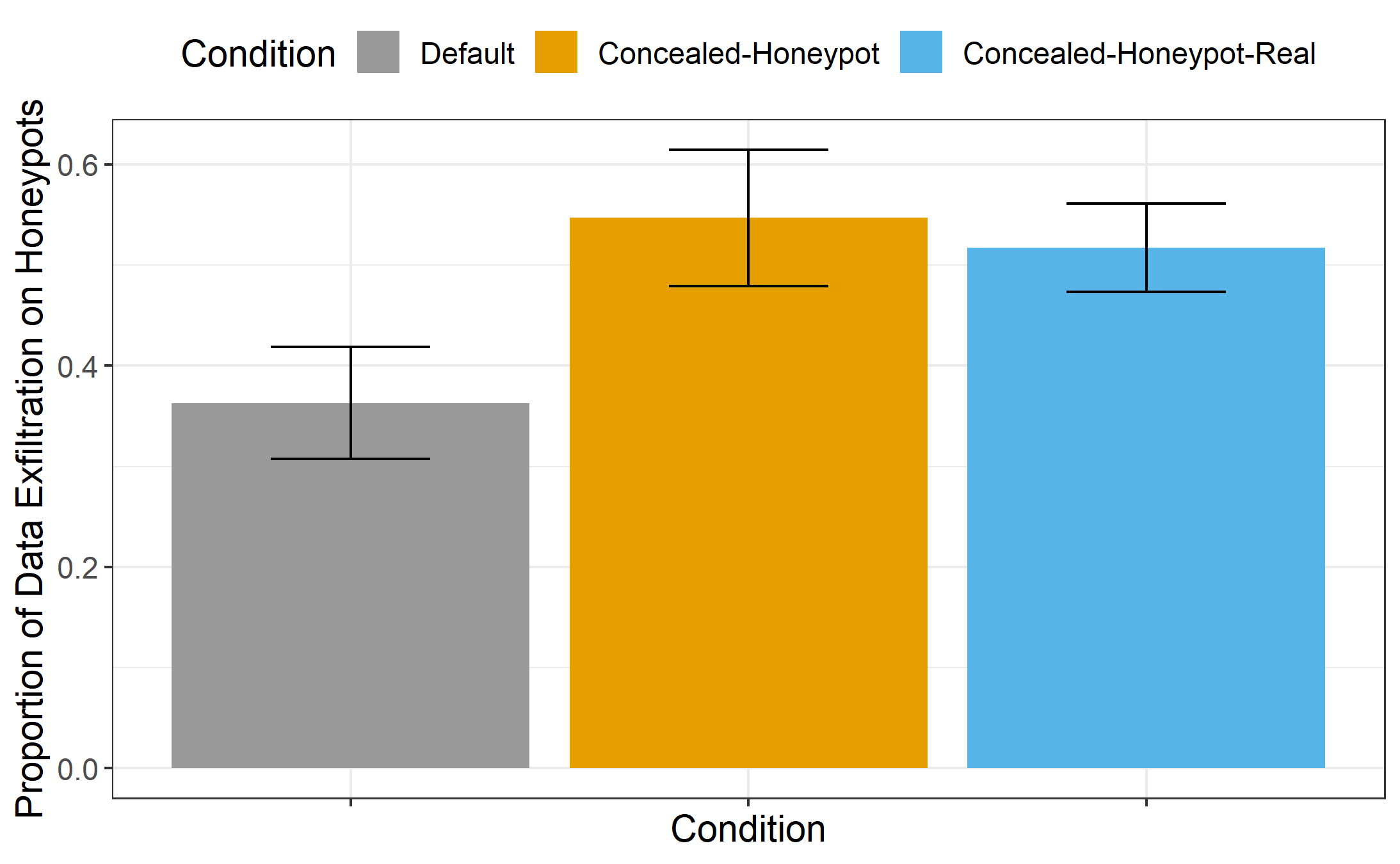}
\caption{Proportion of Data exfiltration from Honeypots}
\label{fig:honeypot_fetch_proportion}
\end{center}
\end{figure}

After successful exploitation of any system, data exfiltration is the next step in the cyber kill chain cycle and shows the attacker's lateral movement in the network. In the HackIT task, we provided a goal to steal the "pin.txt" file from the exploited system. Figure \ref{fig:honeypot_fetch_proportion} shows the proportion of data exfiltration attacks on honeypots. We observe that data exfiltration is higher in concealed honeypot and concealed-honeypot-real conditions compared to the default configuration condition. 

There is a statistically significant difference between groups as determined by one-way ANOVA (\textit{F}(2,44) = 3.05, \textit{p} = 0.053). A Tukey post hoc test revealed that the proportion of data exfiltration is significantly higher in concealed-honeypot condition(\textit{p} = 0.07) and concealed-Honeypot-Real (\textit{p} = 0.10) condition compared to the default condition. There is no statistically significant difference between the concealed-honeypot and concealed-honeypot-real (\textit{p} = 0.91). Our results suggest that attackers steal more data, i.e., use SCP commands more often on concealed conditions compared to the default condition.
\subsection{Score}
\begin{figure}[!htb]
\begin{center}
\includegraphics[scale=0.40]{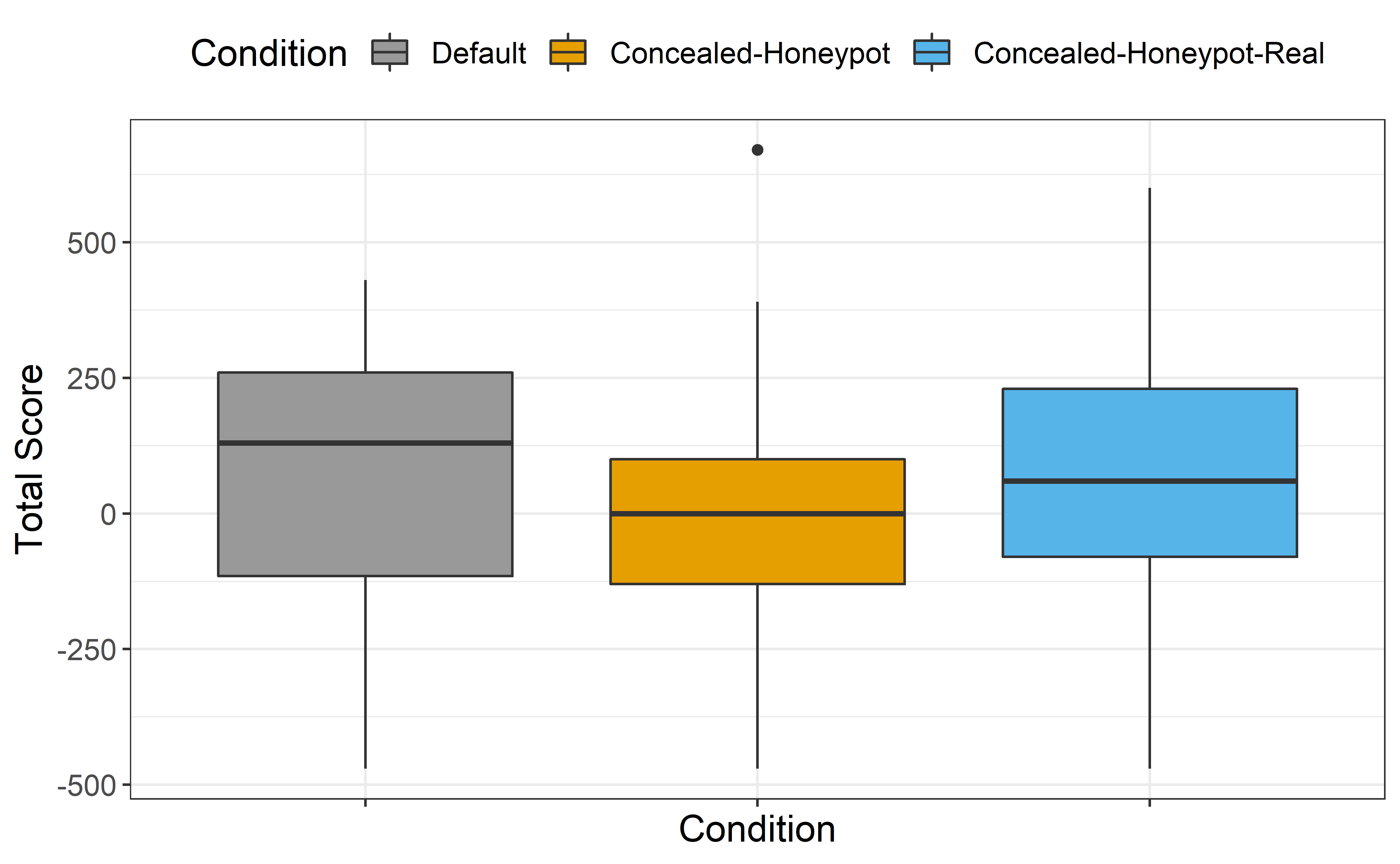}
\caption{Total Score in three conditions}
\label{fig:score}
\end{center}
\end{figure}

Another measure of success of deception is the average score of participants in the three conditions. The score consists of both the actions in the exploit stage and actions in the data exfiltration stage. Figure \ref{fig:score} shows the average score of participants in the three conditions. Participants gain more points in the default condition compared to the conceal-honeypot and conceal-honeypot-real conditions. However, this difference was not statistically significant as determined by one-way ANOVA (\textit{F}(2,46) = 0.216, \textit{p} = 0.89). That is, in terms of the points obtained in the game, there is no difference between conditions.

\subsection{Time Spent on Real vs Honeypots Nodes}

We analyzed the average time the participants spent on each system in the exploit phase and the data exfiltration phase. Time spent in the exploit phase is calculated by the total time spent between the attacker's first entry to the system or the last scp command to the system exploitation command. The time in the data exfiltration phase is calculated from the system exploitation command to the file exfiltration command. 

\begin{figure}[!htb]
\begin{center}
\includegraphics[scale=0.50]{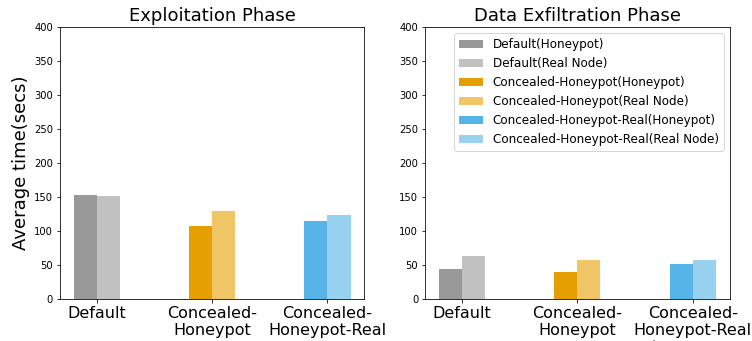}
\caption{Average Time Usage per Target in Pre-Exploit and Exploit phase and Post-Exploit phase}
\label{fig:time1}
\end{center}
\end{figure}

We observed that participants spent similar time on honeypots and real nodes in all conditions as shown in Figure \ref{fig:time1} for both exploitation and data exfiltration phases. One likely reason for this finding is that participants were under time pressure and it is possible that participants are rushing through the network instead of making informed exploit/exfiltrate decisions after full exploration. It can also be the case that the participants can not differentiate between different types of nodes and start to attack quickly and randomly after the first few exploits.

\begin{figure}[!htb]
\begin{center}
\includegraphics[scale=0.50]{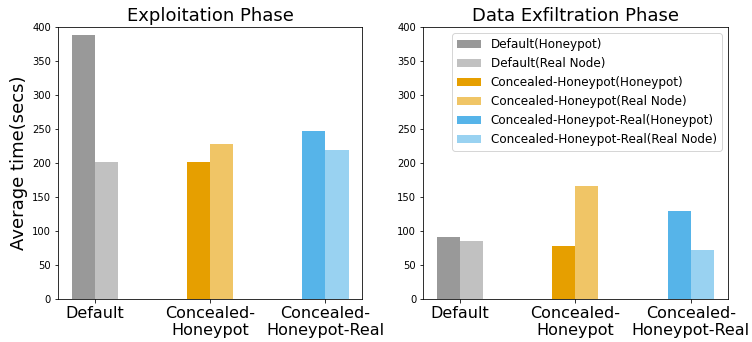}
\caption{Average time spent per Target in exploitation phase and data exfiltration phase during first two exploits}
\label{fig:score2}
\end{center}
\end{figure}

We also analyzed the time spent by participants in their first two exploitation attempts (shown in Figure \ref{fig:score2}). The average time usage of the first two attacks is larger compared to the overall average time usage, suggesting that the participants tend to spend less time on the latter exploits. We observed that participants spent more time on honeypots compared to real nodes in the default condition during the exploitation phase. The time spent on the honeypot and real node was similar in concealed-honeypot and concealed-honeypot-real conditions. During the data exfiltration phase, participants spent more time on real nodes. This could be because both real and honeypots are indistinguishable. In the concealed-honeypot-real nodes, participants spent more time on honeypots where the honeypots were configured to show the features of a real node. 

\subsection{Post-Survey Question Analysis}

After the experiment, we provided a post-survey questionnaire to participants to report their decision making strategies in the task (appendix B). More than half of the participants claimed that their choice of attack targets was affected by the disclosure date of exploits and the estimation from the ``checkHS'' command. Around 10\% participants considered the round-trip time information given by the ``-rtt'' option of ``nmap'' command. 

Furthermore, participants reported the factors that affected the exfiltration decisions were more diverse compared to the features that affected the exploitation decisions. In the options, participants were provided both pre-exploit and exploit features and post-exploit features. Participants reported that the most effective features that affected their exfiltration decision include the ``checkVM'' command that detects a virtual environment, ``checkHS'' estimation, and the number of ``access deny'' encountered during post-exploit reconnaissance. 

In the experiment, we provided a few costly commands to the participants such as ``rtt'' and ``checkHS''. Participants reported that the cost for the commands was not worth. However, in the actual experiment, participants used those commands quite frequently, which represents that despite the cost, participants found those commands useful for their attack decisions. We also asked participants the likely reasons for not fetching the credential file after exploiting a node. Participants reported that they could not locate the credentials file or they did not have time to fetch the file. 

\section{Discussion}
Most often honeypots are deployed without changing the default features such as common ports and running processes. Attackers could easily detect such honeypots and take advantage of such situation to exploit the network. In this paper, we conduct a human experiment using an experimental testbed (HackIT) to study the effectiveness of 2-sided deception, i.e., the effect of concealing the features of both honeypots and real machines against 1-sided deception (concealing the features of honeypot only). We compare these two forms of deception with a default condition, using the default configuration of honeypots in a network. 

Consistent with our hypothesis, the conditions with higher confusion level, i.e., 1-sided deception (concealed honeypots) and 2-sided deception (concealed real and honeypot) conditions, are shown to be more effective in attaining more honeypot attacks and data exfiltration attacks on honeypots compared to the default configuration (no-deception). However, the 1-sided and 2-sided conditions were not significantly different. While making the exploit decisions, the attacks used various commands that provide information about different features in the network. We analyzed the post-survey to understand which features influence the decisions of attackers. Participants reported that the disclosure date of an exploit and the honeypot score influenced their exploitation decision. Furthermore, at the data exfiltration stage, participants reported that indicators such as access denied, empty folders, and deployment of the honeypot on a physical or virtual machine were the most important features. In the future, we will do a more detailed analysis on understanding how various features influence the decisions of attackers in various conditions. 

Although we tried to replicate naturalistic settings, still we made several assumptions on the feature selection, concealing, the cost of feature modification, and availability of tools for attackers. In addition to the experiment design, we also have a limited number of participants which impact the significance of some results. Yet, our results provide some insights to improve the current state-of-the-art techniques of cyberdeception for cyberdefense. Preliminary analysis of our dataset suggests that attackers could detect the honeypots when the default features are used. Deployment of such honeypots may hurt the existing defense system of the network. To make honeypots indistinguishable, defenders could create confusion by making honeypots look like real machines or making real machines look like honeypots, or both. Our result has shown the effectiveness of concealing features on honeypot and real machines over the default configurations. Another limitation of our study is the participants are not red teams or penetration testers. Research in \cite{ferguson2021examining} hired red teams to study cyberdeception, and the decision-making of red teams may impact the results in this study. 

We will continue elaborating on this experimental approach by running laboratory experiments that address the various limitations of the current study. In the near future, we plan to develop a cognitive model that replicates the attacker's behavior in this study. Using Instance-Based Learning Theory \cite{gonzalez2003instance}, we can generate a computational representation of the attacker's decisions. Such models could be used to generate large amounts of synthetic data to test the effectiveness of various deception strategies.

\section*{Acknowledgments}
This research was sponsored by the Combat Capabilities Development Command, Army Research Laboratory and was accomplished under Cooperative Agreement Number W911NF-13-2-0045 (ARL Cyber Security CRA) and by the Army Research Office and accomplished under grant number W911NF-17-1-0370 (MURI Cyberdeception). The views and conclusions contained in this document are those of the authors and should not be interpreted as representing the official policies, either expressed or implied, of the Combat Capabilities Development Command, Army Research Laboratory, or the U.S. Government. The U.S. Government is authorized to reproduce and distribute reprints for government purposes not withstanding any copyright notation here on.

\appendix



\section{Instructions}
Welcome!

This study consists of a hacking task, where you will be playing the role of a hacker. Hackers are people knowledgeable about computers and they use this knowledge to steal data and private information in networks and damage systems that are important for organizations. In this task, your goal is to steal files called "pin.txt" containing credit card pin numbers from different web server computers in a network.

You will be interacting with a network of 40 systems. Your goal is to exploit as many systems as possible to steal as many pin.txt files as possible (these pin.txt files contain credit-card information and pin numbers). You will be rewarded 30 points for getting access into a system and 70 more points for fetching the \textit{pin.txt} file. However, since the \textit{pin.txt} files are highly confidential, the defenders of the network may use deception to create uncertainty. Exploitation and file transfer on deceptive systems will cause equivalent penalty. Please be cautious when choosing attack targets and making the decision to steal \textit{pin.txt} file.

To steal a pin.txt files, you may first probe different systems. Probing means that you try to collect information about the open ports, and the different services running on different web-servers. To find what web-servers are available to you to probe, you may first run \textit{nmap -sL all} command. This command will show you the systems which are available for probing. Now, to probe a system, you need to run the nmap command the following command:

- \textit{nmap System1 [-rtt]}

The output of the nmap command on System1 will also provide a list of vulnerabilities present on the corresponding ports, OS, and vulnerabilities. To know about the benchmark round-trip-time of production systems and the RTT of System1, use option `-rtt`. This will cost you approximately 10 seconds. Before attacking a server, you may use “checkHS” command to detect the existence of deceptive strategies. The reliability of that command is only 50\% and would cost you approximately 10 seconds . It will generate a score from 0 to 1 representing the likelihood of being a deceptive system (0 to 1 as “impossible” to “absolute”). You may use \textit{checkHS System1}  command to get the score. 

To find a most promising attack path, you can collect more detailed information, including its disclosure date, about presented exploits using \textit{info\_ exploit exploitA} command. To exploit a vulnerability on a server, you need to execute the \textit{use\_exploit} command. The  \textit{use\_exploit} command accepts two parameter: the vulnerability to be attacked and the target system name. For example, you may run the following command to exploit \textit{sql\_injection} vulnerability on System1:

- \textit{use\_exploit sql\_injection System1}

Where \textit{sql-injection} is the vulnerability to be exploited and \textit{System1} is the web-server where this vulnerability exists.
The \textit{use\_exploit} command might require a few trials before success. You can also switch to another vulnerability on the same system or another system. Once succeed, you gain access into the server and you may do further reconnaissance with command \textit{ls, ps} and \textit{checkVM}

- \textit{ls, cd [absolute path or relative path]}

The \textit{ls} command lists files present in current directory. The \textit{cd} command helps to traverse through the directories. 

- \textit{ps -A}

The \textit{ps} command lists all running processes on the server.

- \textit{checkVM}

The \textit{checkVM} command simulates virtual environment tools. Its output provide a judgement whether the server is running in suspicious virtual environment. 

After collecting enough information, you can decide whether to steal the file or not. If you decide to steal the file, you may use the scp command to transfer the “pin.txt” file between your system and the remote web-server, which you have exploited. The scp command works in the following way:

- \textit{scp pin.txt 172.22.31.31}

In the command above, \textit{pin.txt} is the credit-card pin file and \textit{172.22.31.31} is your local machine’s IP address. Please note down this IP address \textit{172.22.31.31} as you will have to use it again-and-again to copy pin.txt files from different web-servers.
Once you have copied the file from the exploited web-server, you will get feedback on your current performance in the task. 
You can also choose to logout and terminate the attack anytime you like. After which you may go on to probe, exploit and steal files from other servers.

- \textit{logout}

The flow of the task is also summarized in Figure \ref{fig:HackITTask}. You may use the \textit{help} command any time in the task to list the available commands.

\section{Post Survey Questions}
\begin{itemize}
    \item Q1. Pick the factors that affect your decision on attack target (select all eligible options):
    \begin{itemize}
        \item Operating system
        \item Disclosure date of exploits
        \item \textit{checkHS} estimation
        \item Round trip time measurement
        \item Other
    \end{itemize}
    \item  Q2. If other, please let us know what else affected your decision.
    \item  Q3. Pick the factors that affect your decision on whether to fetch pin.txt file (select all eligible options):
    \begin{itemize}
        \item Operating system
        \item Disclosure date of exploits
        \item \textit{checkHS} estimation
        \item Round trip time measurement
        \item Number of trials to get the foothold
        \item Empty directories encountered
        \item Access of directories denied
        \item Virtual environment detected by \textit{checkVM}
        \item Number of running process
        \item Other
    \end{itemize}
    \item  Q4. If other, please let us know what else affected your decision.
    \item  Q5. Do you think the option \textit{-rtt} of \textit{nmap} command worth the time it takes? If not, pick the largest acceptable time cost:
    
    \begin{itemize*}
        \item 0, 
        \item 2, 
        \item 4, 
        \item 8, 
    \end{itemize*}
    
    \item  Q6. Do you think \textit{checkHS} command worth the time it takes? If not, pick the largest acceptable time cost:
    
    \begin{itemize*}
        \item 0, 
        \item 2, 
        \item 4, 
        \item 8, 
    \end{itemize*}
    
    \item  Q7. Pick the exploits with date that you find too obsolete:
    \begin{itemize}
        \item   HTTP/2 slow read, 2020-03-01
        \item 	LDAPS buffer overflow, 2017-04-11
        \item 	Java deserialize remote code execution, 2015-09-29
        \item 	Remote authentication, 2012-10-23
        \item 	DoS attack, 2010-01-26
        \item 	ASUS remote code execution,	2008-03-25
        \item 	Authentication bypass, 	2006-05-15
        \item 	Remote buffer overflow, 2001-04-04
    \end{itemize}
    \item  Q8. Have you ever given up fetching pin.txt file with the system exploited? If so, pick the options that best describe your reason.
    \begin{itemize}
        \item Can’t locate pin.txt file
        \item I had no time to fetch pin.txt
        \item I discover that the system is honeypot.
        \item To avoid further penalty
    \end{itemize}
    \item  Q9. Please provide us with your feedback about the experiment (optional).
\end{itemize}

\bibliographystyle{named}
\bibliography{ijcai21}

\end{document}